\documentclass[letterpaper,12pt]{article}
\usepackage{color,amssymb,enumerate,float,amsmath}
\usepackage{ifpdf}
\ifpdf
	\usepackage[pdftex]{graphicx}
	\usepackage[pdftex,unicode,implicit]{hyperref}

	\hypersetup{
  	pdftitle     = {}, 
  	pdfkeywords  = {},
  	pdfauthor    = {},
  	pdfcreator   = {pdf\LaTeXe\ with package \flqq hyperref\frqq},
  	pdfproducer  = {pdf\LaTeXe\ with package \flqq hyperref\frqq},
  	pdfpagemode  = UseNone,  
  	pdffitwindow = true,  
  	unicode      = true,
  	plainpages   = true,
  	colorlinks   = true,  
  	citecolor    = black,  
  	urlcolor     = blue, 
  	linkcolor    = black
	}

\else

  \usepackage[dvips]{graphicx}

\fi

\makeatletter
\@addtoreset{equation}{section}
\makeatother

\begin{document}
	
\thispagestyle{empty}

\begin{center}
{\bf \LARGE Batalin-Fradkin-Vilkovisky Quantization of Quadratic Gravity}
\vspace*{15mm}

{\large Jorge Bellorin}$^{1,a}$,
{\large Claudio B\'orquez$^{2,b}$}
{\large and Byron Droguett}$^{1,c}$
\vspace{3ex}

$^1${\it Department of Physics, Universidad de Antofagasta, 1240000 Antofagasta, Chile.}

$^2${\it Facultad de Ingenier\'ia, Universidad San Sebasti\'an, Lago Panguipulli 1390, Puerto Montt, Chile.}

\vspace*{2ex}
$^a${\tt jorge.bellorin@uantof.cl,} \hspace{1em}
$^b${\tt claudio.borquez@uss.cl,} \hspace{1em}
$^c${\tt byron.droguett@uantof.cl}

\vspace*{15mm}
{\bf Abstract}
\begin{quotation}{\small\noindent
We present the Batalin-Fradkin-Vilkovisky quantization of the quadratic gravity theory, which is the most general theory with terms up to quadratic order in curvature. This approach of quantization is based on the Hamiltonian formulation. In this sense, this study contributes to the consistency of the quantum formulation of the theory. With this scheme of quantization we may introduce a broad class of additional conditions on the field variables, by including Lagrange multipliers and time derivatives. We find that a mandatory condition for the validity of the Hamiltonian formulation, previously known from classical analysis, can be incorporated consistently in this quantization. We obtain the propagators of the fields, including the propagators associated with the quantum states of negative norm. The spectrum of masses coincides with the results of Stelle, but distributed on a different way among the fields.
}
\end{quotation}
\end{center}

\newpage
\section{Introduction}
The Hamiltonian is a necessary ingredient in any quantum system. It is the operator that determines the time evolution, according to the postulates of quantum mechanics. Additionally, the process of canonical quantization is based on the existence of conjugated pairs that satisfy canonical commutation relations. In many quantum field theories, one usually deals with the path integral in terms of the Lagrangian. But, for consistency, this approach assumes the existence of a Hamiltonian formulation, which is supposed to have been integrated on canonical momenta, such that the Lagrangian in use is reproduced. In some cases the mathematics of the field theories may be sufficiently complicated such that having an underlying Hamiltonian is not obvious, despite the fact that the majority of the quantum analyses are done on the basis of the Lagrangian path integral. Even when the classical Hamiltonian formulation is known, the quantization may be delicate due to constraints, issues of gauge fixing, the presence of unphysical degrees of freedom, or other factors\footnote{For example, in nonprojectable Ho\v{r}ava gravity \cite{Horava:2009uw,Blas:2009qj}, there are second-class constraints that invalidate the usual Faddeev-Popov quantization based on the Lagrangian, due to the contribution of these constraints to the quantum measure \cite{Bellorin:2019gsc,BellorinII}.}. Therefore, it is important to go back to the basics of the field theory and to perform analysis of the definition of the Hamiltonian formulation and its quantization. In particular, Hamiltonian analyses are very appropriate for characterizing the physical degrees of freedom.

In the context of quantum gravity, the quadratic gravity theory, which we take as the theory defined by having the most general Lagrangian with terms up to quadratic order in curvature (discarding topological contributions), has been widely studied. It has been known for long \cite{Stelle:1976gc} that this theory contains inconsistent modes in its spectrum since they are quantum states with negative norm. On the other hand, quadratic gravity is a renormalizable theory \cite{Stelle:1976gc}, in contrast to the canonical quantization of general relativity (under the perturbative approach) \cite{tHooft:1974toh,Goroff:1985sz,Goroff:1985th,Deser:1974cz,Deser:1974xq}. This feature makes the quadratic gravity an interesting theory to study, since the ultimate theory that reconciles general relativity with quantum mechanics has not been found, and it remains as an essential open question in theoretical physics. One may think that theories with higher orders in curvature could be the answer. Indeed, counterterms of quantum general relativity are terms of higher order in curvature. In this sense, corrections of higher order in curvature are natural in the quantization of the gravitational interaction. Even in the case that the higher orders in curvature do not give the final fundamental theory, they arise as low-energy effective theories of other theories like string theory. Hence, it is interesting to study their properties even in this scenario. For these reasons, it is worth to keep studying the quantum properties of theories of higher order in curvature.

Many of the analyses that have been done on quadratic gravity have been devoted to a better understanding of the quantum modes of its spectrum. We can mention some of them. The negative sign arises in the propagator of certain massive modes, and this is interpreted as a negative norm for the one-particle states representing this propagator \cite{Stelle:1976gc}. The author of Ref.~\cite{Antoniadis:1986tu} proposed using dressed propagators, instead of the bare propagators obtained from the usual canonical quantization, due to the presence of these negative-norm states. The model with only the $R^2$ term in the Lagrangian exhibits interesting properties on its physical degrees of freedom: while the full theory propagates tensor and scalar modes, its perturbation around Minkowski spacetime propagates no tensor mode \cite{Alvarez-Gaume:2015rwa,Hell:2023mph,Barker:2025gon}. There is a special case of values for the parameters of quadratic gravity, called critical gravity \cite{Lu:2011zk}, where a change of the physical features of the propagating modes takes place. This behavior has been also found in a model with only quadratic terms in the Lagrangian \cite{Edery:2019bsh}, i.e, without the Einstein-Hilbert term and the cosmological constant. Propagators are sensitive to gauge-fixing conditions. In this sense, a general class of gauge-fixing conditions in quadratic gravity is studied in Ref.~\cite{Bartoli:1998gs}. Scattering processes of matter particles coupled to quadratic gravity that involve the exchange of the negative-norm states were studied in Ref.~\cite{Salvio:2018kwh}. A different kind of coupling in this theory was analyzed in Ref.~\cite{Hindawi:1995an}. The problem of potential violations of causality in gravity with vertices corrected by higher-order derivatives \cite{Camanho:2014apa}, a kind of analysis that is based on the scattering of gravitons, has gained a great interest. The authors of Ref.~\cite{Edelstein:2021jyu} showed that quadratic gravity is causal, according to the criteria introduced in Ref.~\cite{Camanho:2014apa}. Solutions of the classical field equations of quadratic gravity have also been analyzed, in some cases adopting perturbative or asymptotic limits, in other cases taking subcases of quadratic gravity, as the case of Weyl gravity \cite{Stelle:1977ry,Boulware:1983td,Teyssandier:1989dw,Lu:2015psa,Edery:2018jyp,Bellorin:2025kir}.

Analyses devoted to the quantization of quadratic gravity in terms of the underlying Hamiltonian formulation are fewer in number. The classical Hamiltonian of this theory is known \cite{Buchbinder:1987vp}; therefore, there is an initial step\footnote{The Hamiltonian formulation of quadratic gravity was previously analyzed in Ref.~\cite{Boulware:1983yj}. However, the rigorous proof of the validity of the Hamiltonian based on the equivalence between the Lagrangian and the Hamiltonian equations of motion was presented in Ref.~\cite{Buchbinder:1987vp}. Another previous analysis on quadratic gravity can be found in Ref.~\cite{Kaku:1982xt}, where the Hamiltonian formulation of the Weyl gravity theory was studied.}. In this study, a form of quantization was presented, using the formalism of the path integral developed by Faddeev \cite{Faddeev:1969su}. This formalism only admits gauge-fixing conditions that are of canonical type. A proposal for a scheme of quantization, using the Hamiltonian formalism, that can include the negative-norm states in the physical spectrum is given in Ref.~\cite{Salvio:2024joi}. Further advances in the Hamiltonian formulation of quadratic gravity, specially for the case of Weyl gravity, can be found in Ref.~\cite{Kluson:2013hza}.

The Batalin-Fradkin-Vilkovisky (BFV) formalism \cite{Fradkin:1975cq,Batalin:1977pb,Fradkin:1977xi} was developed to undertake the quantization of relativistic systems with constraints defined in terms of the Hamiltonian formulation, in such a way that covariant gauges can be implemented in the quantization and the unitarity of the theories can be scrutinized. This is a very important approach, since the quantization of systems with first-class constraints using the recipe of the Faddeev path integral is limited to a certain class of gauge-fixing conditions: those conditions that are canonical. This is the case of the quantization presented in Ref.~\cite{Buchbinder:1987vp}, as we have commented. Therefore, the BFV formalism allow us to combine the quantization in terms of the underlying Hamiltonian with a broad class of gauge-fixing conditions, including the covariant ones for the case of relativistic theories. This quantization was implemented in general relativity \cite{Fradkin:1975sj}, with the aim of establishing its unitarity under covariant gauges.

There is a technical aspect in the Hamiltonian formulation of quadratic gravity that affects the possible additional conditions on the field variables one can impose. The Hamiltonian equations of motion of this theory were presented explicitly in Ref.~\cite{Bellorin:2025kir}. A consistency condition was found in this study: the condition that the (perturbative) spatial metric must be traceless. This is necessary in order for the Hamiltonian formulation be equivalent to the original covariant formulation, since the equivalence must hold at the level of the classical equations of motion. Since the quantization is based on the approach of the canonical quantization, this is a required condition for the quantum theory.

In this paper we study the BFV quantization of quadratic gravity, taking the classical Hamiltonian formulation developed by Buchbinder and Lyahovich \cite{Buchbinder:1987vp}. We have two main goals: first, we want to study the complete formulation of the BFV path integral, which may serve as a master formula for a variety of quantum analysis. The BFV formalism allows us to introduce gauge-fixing conditions that depend on time derivatives and Lagrange multipliers. This is required in special for covariant gauges. Parallel to this, we want to understand the role of the mandatory additional condition on the field variables that we have commented, in this case at the level of the quantum theory. Second, we study the quantum modes of the spectrum on the basis of the BFV quantization with the appropriate gauge-fixing condition. We study the several propagators in order to elucidate how the inconsistent modes arise in this Hamiltonian formalism.

This paper is organized as follows. In section 2 we present the BFV quantization of the theory of quadratic gravity. In section 3 we perform the perturbative version at the linearized order, introducing the specific gauge-fixing condition, and we obtain the propagators. We present some conclusions, and, in the appendix we show the propagators of all the quantum fields obtained under this quantization.


\section{BFV phase space, BRST charge, and path integral}
Quadratic gravity is given by the covariant Lagrangian\footnote{We put an upper label $^{(4)}$ on spacetime tensors.}
\begin{equation}
	\mathcal{L} = 
	\sqrt{-g^{\text{\tiny(4)}}} \left( \kappa^{-2} R^{\text{\tiny(4)}} 
	- \alpha R^{\text{\tiny(4)}}_{\mu\nu} R^{\text{\tiny(4)}}{}^{\mu\nu}
	+ \beta {R^{\text{\tiny(4)}}}^2 \right) \,,
	\label{LagrangianR2}
\end{equation}
where $\kappa^{-2}$, $\alpha$ and $\beta$ are arbitrary coupling constants. The quantization of this theory was studied in \cite{Stelle:1976gc}, with the aim of proving its renormalization. The starting point is that (\ref{LagrangianR2}) gives the most general action for a theory with terms up to quadratic order in the curvature tensor, discarding topological contributions (in our case we do not consider the cosmological constant). The rigorous Hamiltonian formulation of this theory was presented in \cite{Buchbinder:1987vp}, based on the equivalence between the Lagrangian and Hamiltonian equations of motion. The method presented in \cite{Buchbinder:1987vp} is a generalization of Ostrogradsky's method for the Hamiltonian formulation of systems of higher order in time derivatives, including constraints \cite{Gitman}. As in general relativity, the Arnowitt-Deser-Misner (ADM) variables $N,N^i,g_{ij}$ are the appropriate ones for the Hamiltonian formulation. For quadratic gravity, one of the main steps is that one uses the extrinsic curvature tensor $K_{ij}$, together with its canonically conjugate momentum, as the additional canonical variables associated with the higher order in time derivatives. Thus, the phase space of the classical theory is spanned by the canonical pairs $\{ (g_{ij}, \pi^{ij}) , (K_{ij}, P^{ij}) \}$. It is convenient to introduce a nonrelativistic index $A=0,i$. With this index, for example, $N$ and $N^i$ can be denoted as a single object in the way $N^A = N, N^i$. In quadratic gravity, the four $N^A$ are Lagrange multipliers, as in general relativity. We list the several combinations we use in the Hamiltonian formulation: 
\begin{equation}
\begin{split}
&
\upsilon_m = \alpha - m\beta \,, \quad  m=1,2,3\ldots,
\\&
\gamma = ( 8 \alpha \upsilon_3 )^{-1} \,,
\\&
V_i = \nabla^j K_{ij} - \nabla_i K  \,,
\\&
L = K_{ij} K^{ij} - K^2 \,,
\quad K = g^{ij} K_{ij} \,,
\\&
J_{ij} = 2 K_{ik} K_{j}{}^k - K_{ij} K \,,
\\&  
G^{ijkl} =  
-\alpha ( g^{ik} g^{jl} + g^{il} g^{jk} )
- 2 \upsilon_4 g^{ij} g^{kl} \,,
\\&
{G}^{-1}_{ijkl} =
-(4\alpha)^{-1} ( g_{ik} g_{jl} + g_{il} g_{jk} )
+ \gamma\upsilon_4 g_{ij} g_{kl} \,,
\\&
Q^{i j} = 
P^{i j} - 2 \sqrt{g} \left( - \alpha  R^{ij} + g^{ij} ( \kappa^{-2} - \upsilon_2 L + 2 \beta R ) \right)  
\,, 
\quad
\tilde{Q}_{ij} = G^{-1}_{ijkl} Q^{kl}  \,,
\\&
V =  
\sqrt{g} \left( 
\kappa^{-2} ( L + R ) + \alpha ( L^2 + R_{ij} R^{ij} - 2 V_i V^i ) - \beta ( L + R )^2 \right) 
\,.
\end{split}
\end{equation}
In the Hamiltonian version that we study, two conditions on the coupling constants are imposed: $\alpha \neq 0$ and $\upsilon_3 \neq 0$. The bulk part of the Hamiltonian is a sum of constraints, such that the classical canonical action is
\begin{equation}
S =
\int\! dt d^3x \left(
\pi^{ij} \dot{g}_{ij} + P^{ij} \dot{K}_{ij} 
- N^A T_A 
\right) \,,
\label{canonicalaction}
\end{equation}
where
\begin{eqnarray}
&&
T_0 = 
\frac{1}{2\sqrt{g}} \tilde{Q}_{ij} Q^{ij}
+ \left( \nabla_{ij} + J_{ij} \right) P^{ij} 
+ 2 K_{ij} \pi^{ij} + V \,,
\label{T0}
\\&& 
T_i =
- 2 g_{ij} \nabla_k \pi^{jk} 
- 2 \nabla_k ( K_{ij} P^{jk} ) 
+ \nabla_i K_{jk} P^{jk}  \,.
\label{Ti}
\end{eqnarray}
The four $T_A$ are first-class constraints, $T_A = 0$. Their algebra corresponds to the algebra of spacetime diffeomorphisms, which is given by
\begin{equation}
\begin{split}
&
{\textstyle 
	\left\{ \int\! \omega^{i} T_{i} \,, \int\! \eta^{j} T_{j}  \right\}
	=
	\int\! \left( \omega^{i} \partial_{i} \eta^{j}
	- \eta^{i} \partial_{i} \omega^{j} \right) T_{j} } \,,
\\&                 
{\textstyle 
	\left\{ \int\! \omega^{i} T_{i} \,, \int\! \sigma T_{0}  \right\}
	=
	\int\! \omega^{i} \partial_i \sigma T_0 } \,,
\\&
{\textstyle 
	\left\{ \int\! \rho T_0  \,, \int\! \sigma T_{0}  \right \} =
	\int\! ( \rho \partial^i \sigma - \sigma \partial^i \rho ) T_i } \,.
\end{split}
\label{algebra}
\end{equation}
The physical degrees of freedom in the phase space are equal to 16, corresponding to 8 independent physical modes.

There is a technical detail necessary to complete the validity of this Hamiltonian formulation of quadratic gravity. The theorem of Buchbinder and Lyahovich \cite{Buchbinder:1987vp} establishes the validity of the Hamiltonian based on the equivalence of the equations of motion with the original Lagrangian ones. When first-class constraints are present, they induce arbitrary functions that are not fixed by the equations of motion. Therefore they must be fixed for all time by imposing additional conditions. This is equivalent to the gauge-fixing procedure in theories with gauge symmetries using the usual covariant Lagrangian. In the case of quadratic gravity, when comparing the equations of motion with the covariant ones, if $\kappa^{-2} \neq 0$, it turns out that one specific additional condition must be imposed in order to have a complete equivalence between both formulations. Notice that this comparison must be done a the level of the classical theory, since the quantum theory is obtained from a canonical quantization procedure; hence we need to start with a consistent classical Hamiltonian. We will discuss the additional condition explicitly in the next section using perturbative variables.

We proceed to the BFV quantization of this theory. We promote the Lagrange multipliers $N^A$ associated with the first-class constraints to the range of canonical variables. Thus, we have the new canonical pairs $(N,P_{N})$ and $(N^{i},\tau_i)$. We can group the new momenta in one single variable: $\pi_A = P_{N},\tau_i$. According to the BFV formalism, we define the functions $G_a= T_A,\pi_A$. In this theory, the functions $G_a$ satisfy an involution relationship. This takes the form of the algebra
\begin{equation}
\{G_a \,, G_b\} = U_{ab}^{c} G_c \,.
\end{equation}
An important technical aspect, which is also present in the canonical quantization of general relativity, is the fact that the structure coefficients $U^c_{ab}$ depend on the metric components. We introduce two canonically conjugate pairs of fermionic ghosts, which we denote by $(\eta_1^{A},\mathcal{P}^{1}_A)$ and $(\eta_2^{A},\mathcal{P}^{2}_A)$. The splitting of these variables is $\eta_1^A= \eta_1, \eta_1^k$,  $\eta_2^A= \eta_2, \eta_2^k$, $\mathcal{P}^1_A= \mathcal{P}^1, \mathcal{P}_k^1 $, $\mathcal{P}^2_A=  \mathcal{P}^2, \mathcal{P}_k^2 $.

Following the general BFV formalism, for this theory the gauge-fixed quantum Hamiltonian is given by 
\begin{equation}
	\mathcal{H}_{\Psi}= \left\lbrace\Psi,\Omega\right\rbrace \,.
\label{generalgfhamiltonian}
\end{equation}
$\Omega$ is the generator of the Becchi-Rouet-Stora-Tyutin (BRST) symmetry. It is defined symbolically by the formula
\begin{equation}
	\Omega = 
	T_A \eta_1^A
	+ \pi_A \eta_2^A
	- \frac{1}{2}U_{AB}^{C}\eta_1^{A}\eta_1^{B}\mathcal{P}^{1}_C \,,
	\label{omega}
\end{equation}
which results, explicitly, as follows:
\begin{equation}
	\Omega =
	\int d^3x \left(
	T_A\eta_1^{A} + \pi_A\eta_2^{A}
	- \eta_1^i\partial_i \eta_1 \mathcal{P}^1
	- \eta_1^i\partial_i\eta_1^j\mathcal{P}_j^1
	\right) \,.
\end{equation}
An essential consistency condition of the whole BFV formalism, which is the basis of the BRST symmetry, is that $\Omega$ must satisfy\footnote{$\Omega$ is an odd variable, therefore (\ref{bfvconditions}) is a nontrivial condition.}
\begin{equation}
	\{ \Omega \,, \Omega\} = 0 \,.
\label{bfvconditions}	
\end{equation} 
We have checked that the BRST generator $\Omega$ satisfies this condition. $\Psi$ is the gauge fermion that implements the preferred additional conditions on the arbitrary field variables.

The definition of the BFV path integral requires to choose a gauge fermion $\Psi$, and the BFV main theorem ensures that the path integral is independent of $\Psi$. The path integral takes the form
\begin{equation}
\begin{split} 
	&
	Z = 
	\int\mathcal{D} \mathcal{V} 
	\exp \left[
	i \int dt d^{3}x
	\Big(
	  \pi^{ij} \dot{g}_{ij} 
    + P^{ij} \dot{K}_{ij}
    + \pi_A \dot{N}^{A}
    + \mathcal{P}_{A}^{1}\dot{\eta}^{A}_{1}
    + \mathcal{P}_{A}^{2}\dot{\eta}^{A}_{2}
    - \mathcal{H}_\Psi\Big) 
    \right] \,,
    \\&
    \mathcal{D}\mathcal{V} =
    \mathcal{D} g_{ij} \mathcal{D} \pi^{ij} 
    \mathcal{D}K_{ij}\mathcal{D}P^{ij}\mathcal{D}N^A \mathcal{D}\pi_A 
	\mathcal{D}\eta^A_{1} \mathcal{D}\mathcal{P}^{1}_A \mathcal{D}\eta^A_{2} \mathcal{D}\mathcal{P}^{2}_A \,.
\end{split}
\end{equation}
Notice that there is no quantum measure associated with the first-class constraints. According to the BFV formalism, an appropriate form of the gauge fermion that includes time derivatives in the additional conditions is
\begin{equation}
 \Psi = \mathcal{P}_A^{1}N^{A} + \mathcal{P}_A^{2}\chi^{A} \,,
\end{equation}
where 
\begin{equation}
 \dot{N}^A - \chi^A = 0
 \label{bfvgaugegeneral}
\end{equation}
are the conditions on the arbitrary functions to be imposed. This procedure requires to fix the four factors $\chi^A = \chi,\chi^i$. We fix the definitive form of $\chi^A$, and consequently of the Hamiltonian and the path integral, in the perturbative approach in the next section.

\section{The linearized theory and the propagators}
\subsection{Linearized version of the theory: the additional conditions}
We make a convenient change of notation on the BFV ghosts, as follows
\begin{equation}
\begin{split}
	&	
	\eta_{1}^{A} = 
	(\eta_{1},\eta_{1}^{i}) = (C,C^{i}) \,,
	\quad
	\eta_{2}^{A} = ( \eta_{2},\eta_{2}^{i} ) = (\mathcal{P},\mathcal{P}^{i}) \,,
	\\&
	\mathcal{P}_{A}^{1} = ( \mathcal{P}^{1},\mathcal{P}^{1}_i ) = (\bar{\mathcal{P}},\bar{\mathcal{P}}_i)\,,
	\quad
	\mathcal{P}_{A}^{2} = ( \mathcal{P}^{2},\mathcal{P}^{2}_i ) = ( \bar{C},\bar{C}_i )\,.
\end{split}
\end{equation}
After integration on $\mathcal{P}, \mathcal{P}^i , \bar{\mathcal{P}} , \bar{\mathcal{P}}_i$, the remaining ghosts $C,C^i, \bar{C} , \bar{C}_i$ become equivalent to the Faddeev-Popov ghosts of the covariant formulation.

A technical detail of the Hamiltonian formulation of quadratic gravity is that a nonvanishing zeroth mode for the momentum $P^{ij}$ is required to define the Minkowski background \cite{Bellorin:2025kir}. Thus, the perturbative variables are defined by 
\begin{eqnarray}
& g_{ij} = \delta_{ij} + h_{ij} \,,
  \quad
 N = 1 + n \,,
\quad
N^i = n^i \,, \quad
P^{ij} = 2\kappa^{-2} \delta^{ij} + p^{ij}  \,,
\end{eqnarray}
and for the rest of fields, we keep the original notation. We use the three-dimensional transverse-longitudinal decomposition for tensors: if $\sigma_{ij}$ is a symmetric spatial tensor, we decompose it in the way
\begin{equation}
\sigma_{ij} = 
\partial_{ij} \Delta^{-1} \sigma^L
+ \partial_{(i} \sigma_{j)}^L
+ \frac{1}{2} \tau_{ij} \sigma^T  
+ \sigma_{ij}^{T} \,,
\label{decomposition}
\end{equation}
where $\Delta = \partial_{kk}$, and with the conditions: $\sigma_{kk}^T = \partial_k \sigma_{kj}^T = \partial_k \sigma_k^L = 0$, and $\tau_{ij}$ is the tranverse projector,
\begin{equation} 
	\tau_{ij} = \delta_{ij} - \partial_{ij} \Delta^{-1}  \,.
\end{equation} 
For a spatial vector $\xi_i$ we use
\begin{equation}
\xi_i =
  \partial_i \xi^L + \xi_i^T \,,
\label{vectors}
\end{equation}
with $\partial_k \xi_k^T = 0$. We use the combinations
\begin{equation}
	\eta = 3\alpha - 8\beta \,,
	\quad
	\tilde{\eta} = 5\alpha - 16\beta \,,
	\quad
	\mu = \frac{\kappa^{-2}}{\alpha} \,.
\end{equation}

As we have commented in the previous section, there is a mandatory additional condition required to hold the validity of the Hamiltonian formulation, which is deduced at the level of the classical theory. One of the linearized equations of motion derived from the canonical action (\ref{canonicalaction}) takes the form \cite{Bellorin:2025kir}
\begin{equation}
	\left( \eta \Box + \kappa^{-2} \right) \Box h^T
	- 2 \left( \upsilon_4 \Box + \kappa^{-2} \right) \phi 
	= 
	- 2\kappa^{-2} \Delta ( h^L + h^T ) \,,
	\label{eompreaaditional}
\end{equation}
where
\begin{equation}
	\Box = -\partial_{00} + \Delta \,,
	\quad
	\phi = \ddot{h}^L - 2 \Delta ( \dot{n}^L + n ) \,.
\end{equation}
On the other hand, the linearized covariant field equation deduced from the Lagrangian (\ref{LagrangianR2}) is equal to (\ref{eompreaaditional}), but with the right-hand side equal to zero. Therefore, the equivalence of the classical Hamiltonian formulation with the original Lagrangian one, in the case $\kappa^{-2} \neq 0$, only holds if one imposes the additional condition $ \Delta ( h^L + h^T )  = 0 $. Since in the whole formalism we are assuming spatial boundary such that $\Delta$ is invertible, this condition is equivalent to
\begin{equation}
 h^L = - h^T \,,
 \label{traceless}
\end{equation}
which is equivalent to saying that the linear-order spatial metric is traceless.

Now, we may define the general additional conditions on the arbitrary functions. We fix the factors $\chi^A$ following several rules. They are linear on the fields. They are based on the functional condition (\ref{bfvgaugegeneral}), and include no dimensional coupling constants. As a consequence, $\chi^A$ do not depend on time derivatives and the possible combinations of fields arising in them are restricted by dimensionality. The $\chi^A$ are local expressions; hence only the identity and spatial derivatives can operate on the fields. We want the rigid (global) symmetry of spatial rotations to be preserved. We do not consider the dependence of the additional conditions on any ghost fields. Finally, the condition (\ref{traceless}) must be imposed as part of the additional conditions. After these considerations, splitting $\chi^i$ as in (\ref{vectors}), we find the set of additional conditions:
\begin{eqnarray}
 &&
 \chi =
 \sigma_1\Delta n^L+\sigma_2(K^T+K^L) \,,
 \label{gauechi}
 \\&&
 \chi^L = 
 \frac{1}{\epsilon} ( h^L + h^T ) \,,
 \label{chiL}
 \\&&
 \chi_i^T = 
 \sigma_3 \Delta h_i^L \,,
 \label{gaugechiiT}
\end{eqnarray}
where $\sigma_1,\sigma_2,\sigma_3$ and $\epsilon$ are dimensionless constants. We may take the limit $\epsilon \rightarrow 0$, such that the combination of Eqs.~(\ref{chiL}) and (\ref{bfvgaugegeneral}) becomes equivalent to the condition (\ref{traceless}) in that limit.

Given the additional conditions, we may compute explicitly the Hamiltonian (\ref{generalgfhamiltonian}). In the following we express the canonical Lagrangian using the transverse-longitudinal decomposition. Since the expressions are large, we separate the kinetic terms from the Hamiltonian, $ \mathcal{L} = \mathcal{K} - \mathcal{H}_\Psi $. The kinetic terms get an orthogonal decomposition, as follows:
\begin{equation}
\begin{split}
\mathcal{K} =& 
  \pi^L \dot{h}^L
+ \frac{1}{2} \pi^T \dot{h}^T
+ \frac{1}{2} \partial_k \pi_l^L  \partial_k \dot{h}_l^L 
+ \pi^{T}_{ij} \dot{h}_{ij}^{T}
+ p^L \dot{K}^L 
+ \frac{1}{2} p^T \dot{K}^T 
+ \frac{1}{2} \partial_k p_l^L  \partial_k \dot{K}_l^L 
+  p^{T}_{ij} \dot{K}_{ij}^{T}
\\&
+ P_N \dot{n}
+ \partial_k \tau^L \partial_k \dot{n}^L 
+ \tau_k^T \dot{n}_k^T 
+ \bar{\mathcal{P}}\dot{C}
+ \mathcal{P}\dot{\bar{C}}
+ \partial_k \bar{\mathcal{P}}^L \partial_k \dot{C}^L
+ \partial_k \mathcal{P}^L \partial_k \dot{\bar{C}}^L  
+ \bar{\mathcal{P}}_k^T\dot{C}_{k}^{T}
+ \mathcal{P}_k^{T}\dot{\bar{C}}_k^T
\,.
\end{split}
\end{equation}
We present the Hamiltonian separated in sectors,
\begin{equation}
 \mathcal{H}_\Psi = 
   \mathcal{H}_\Psi^{\text{\tiny scalar}}
 + \mathcal{H}_\Psi^{\text{\tiny vector}}
 + \mathcal{H}_\Psi^{\text{\tiny tensor}} \,,
\end{equation}
where
\begin{equation}
\begin{split}
 \mathcal{H}_\Psi^{\text{\tiny scalar}} =&
 - \frac{1}{2}\gamma\upsilon_2 {p^T}^2
 - \frac{1}{2}\gamma\eta {p^L}^2
 + \gamma\upsilon_4 p^T p^L
 - \gamma p^T\left( 2\alpha\beta \Delta + \kappa^{-2}\upsilon_4 \right)h^T
 - \kappa^{-2} \gamma \tilde{\eta} p^L h^L
 \\&
 + \gamma p^L \left( \kappa^{-2}\eta + 2\alpha\upsilon_4\Delta  \right)h^T
 + \kappa^{-2}\gamma\eta p^T h^L
 + \pi^T K^T
 + 2\pi^LK^L
 \\&
 + \gamma h^T\left( 2\alpha^2\upsilon_4\Delta^2
 - \kappa^{-2}\alpha\upsilon_5\Delta
 - \frac{1}{2}\kappa^{-4}\eta \right)h^T
 + \gamma h^T\left( \kappa^{-4} \tilde{\eta} 
 + 2\kappa^{-2}\alpha \upsilon_2\Delta \right)h^L
 \\&
 - \frac{1}{2}\kappa^{-4}\gamma \left( 11\alpha - 32\beta \right) {h^L}^2
 + K^T \left( 2\alpha\Delta + \frac{1}{2} \kappa^{-2} \right) K^T 
 + 2\kappa^{-2} {K^L}^2
 \\&
 - 2\kappa^{-2}K^T K^L
 + p^L \Delta n 
 + 2 \pi^L \Delta n^L 
 + \kappa^{-2} n\Delta h^L
 - 2\kappa^{-2} n^L\Delta ( K^T - K^L )
 \\&
 + \sigma_1 P_{N} \Delta n^L 
 + \sigma_2 P_{N}( K^T + K^L )
 - \epsilon^{-1} \tau^{L}\Delta(h^T+h^L) 
 \\&
 + \bar{\mathcal{P}}\mathcal{P}
 - \bar{\mathcal{P}}^L\Delta\mathcal{P}^L
 + \sigma_2\bar{C}\Delta C
 - 2\epsilon^{-1}\bar{C}^L\Delta^2C^L 
 + \sigma_1\bar{C}\Delta\mathcal{P}^L \,,
\end{split}
\end{equation}
\begin{equation}
\begin{split}
  \mathcal{H}_\Psi^{\text{\tiny vector}} 
 =& 
   \frac{1}{8\alpha} p_i^L \Delta p_i^L
 + \frac{\mu}{2} p_i^{L} \Delta h_i^L  
 - \pi_i^{L} \Delta K_i^L  
 + \frac{1}{2} \kappa^{-2} \mu h_i^L \Delta h_i^L
 - \frac{1}{2} K_i^L\left( \alpha \Delta^2  
 + 3 \kappa^{-2}\Delta \right)K_i^{L}
 \\&
 - \pi_i^{L} \Delta n_i^T
 - 2\kappa^{-2}n_i^T\Delta K_{i}^L
 + \sigma_3 \tau_i^T\Delta h_i^{L}
 + \bar{\mathcal{P}}_i^T\mathcal{P}_i^{T}
 + 2\sigma_3\bar{C}^T_i\Delta C^{T}_i
 \,,
\end{split}
\end{equation}
\begin{equation}
\begin{split}
 \mathcal{H}_\Psi^{\text{\tiny tensor}}
 =& 
 - \frac{1}{4\alpha} p_{ij}^{T}p_{ij}^{T} 
 + p^{T}_{ij}\left( \frac{1}{2}\Delta - \mu
 \right)h_{ij}^{T}
 + 2\pi_{ij}^{T} K_{ij}^{T}
 + \kappa^{-2} h_{ij}^{T}  \left( \frac{3}{4} \Delta - \mu \right) h_{ij}^{T}
 \\&
 + 3\kappa^{-2} K_{ij}^{T} K^{T}_{ij}
 \,.
\end{split}
\label{HamiltonianTensors}
\end{equation}

In order to simplify the field content of the resulting theory, the ghost fields $\mathcal{P}$, $\bar{\mathcal{P}}$, $\mathcal{P}^L$, $\bar{\mathcal{P}}^L$, $\mathcal{P}^T_i$, $\bar{\mathcal{P}}^T_i$ can be integrated directly\footnote{After the integration, the action maintain the Hamiltonian form for all fields that are not the ghosts $C, \bar{C} , C^i, \bar{C}_i$.}. By doing this, we obtain the kinetic terms
\begin{equation}
\begin{split}
	\mathcal{K} =& 
	\pi^L \dot{h}^L
	+ \frac{1}{2} \pi^T \dot{h}^T
	+ \frac{1}{2} \partial_k \pi_l^L  \partial_k \dot{h}_l^L 
	+ \pi^{T}_{ij} \dot{h}_{ij}^{T}
	+ p^L \dot{K}^L 
	+ \frac{1}{2} p^T \dot{K}^T 
	+ \frac{1}{2} \partial_k p_l^L  \partial_k \dot{K}_l^L 
	\\&
	+ p^{T}_{ij} \dot{K}_{ij}^{T}
	+ P_N \dot{n}
	+ \partial_k \tau^L \partial_k \dot{n}^L 
	+ \tau_k^T \dot{n}_k^T 
	+ \dot{C} \dot{\bar{C}}
    + \partial_k \dot{C}^L \partial_k \dot{\bar{C}}^L + \sigma_1 \dot{C}^L \Delta\bar{C}  
	+ \dot{C}_i^T \dot{\bar{C}}_i^T
	\,,
\end{split}
\end{equation}
and the Hamiltonian for the scalar and vector sectors, as follows:
\begin{equation}
	\begin{split}
		\mathcal{H}_\Psi^{\text{\tiny scalar}} =&
		- \frac{1}{2}\gamma\upsilon_2 {p^T}^2
		- \frac{1}{2}\gamma\eta {p^L}^2
		+ \gamma\upsilon_4 p^T p^L
		- \gamma p^T\left( 2\alpha\beta \Delta + \kappa^{-2}\upsilon_4 \right)h^T
		- \kappa^{-2} \gamma \tilde{\eta} p^L h^L
		\\&
		+ \gamma p^L \left( \kappa^{-2}\eta + 2\alpha\upsilon_4\Delta  \right)h^T
		+ \kappa^{-2}\gamma\eta p^T h^L
		+ \pi^T K^T
		+ 2\pi^LK^L
		\\&
		+ \gamma h^T\left( 2\alpha^2\upsilon_4\Delta^2
		- \kappa^{-2}\alpha\upsilon_5\Delta
		- \frac{1}{2}\kappa^{-4}\eta \right)h^T
		+ \gamma h^T\left( \kappa^{-4} \tilde{\eta} 
		+ 2\kappa^{-2}\alpha \upsilon_2\Delta \right)h^L
		\\&
		- \frac{1}{2}\kappa^{-4}\gamma \left(  11\alpha - 32\beta \right) {h^L}^2
		+ K^T \left( 2\alpha\Delta + \frac{1}{2} \kappa^{-2} \right) K^T 
		+ 2\kappa^{-2} {K^L}^2
		\\&
		- 2\kappa^{-2}K^T K^L
		+ p^L \Delta n 
		+ 2 \pi^L \Delta n^L 
		+ \kappa^{-2} n\Delta h^L
		- 2\kappa^{-2} n^L\Delta ( K^T - K^L )
		\\&
		+ \sigma_1 P_{N} \Delta n^L 
		+ \sigma_2 P_{N}( K^T + K^L )
		- \epsilon^{-1} \tau^{L}\Delta(h^T+h^L) 
		\\&
		+ \sigma_2\bar{C}\Delta C
		- 2\epsilon^{-1}\bar{C}^L\Delta^2C^L  \,,
	\end{split}
\end{equation}
\begin{equation}
	\begin{split}
		\mathcal{H}_\Psi^{\text{\tiny vector}} 
		=& 
		\frac{1}{8\alpha} p_i^L \Delta p_i^L
		+ \frac{\mu}{2} p_i^{L} \Delta h_i^L  
		- \pi_i^{L} \Delta K_i^L  
		+ \frac{1}{2} \kappa^{-2} \mu h_i^L \Delta h_i^L
		- \frac{1}{2} K_i^L\left( \alpha \Delta^2  
		+ 3 \kappa^{-2}\Delta \right)K_i^{L}
		\\&
		- \pi_i^{L} \Delta n_i^T
		- 2\kappa^{-2}n_i^T\Delta K_{i}^L
		+ \sigma_3 \tau_i^T\Delta h_i^{L}
		+ 2\sigma_3\bar{C}^T_i\Delta C^{T}_i
		\,.
	\end{split}
\end{equation}
The Hamiltonian (\ref{HamiltonianTensors}) of the tensor sector remains unaltered after the integration.

\subsection{Propagators}
From the previous second-order canonical Lagrangian, we obtain the propagators of all quantum fields. We compute the inverse of the matrix of coefficients in Fourier space. Since in our approach all quantum vector and tensor fields satisfy transverse and traceless conditions, we introduce unit orthogonal basis on these subspaces. For the vectors we introduce a basis $\{ e_i^{x} \}$ on the transverse space, such that $\vec{k}_i e^x_i = 0$ and $e^x_i e^y_i = \delta^{xy}$. The index $x$ runs for the two dimensions of the plane orthogonal to $\vec{k}$. For the tensors we introduce a basis $\{ e^X_{ij} \}$, which is given by symmetric tensors, $e^X_{ij} = e^X_{ji}$, that satisfy $\vec{k}_i e^X_{ij} = 0$, $ e^X_{ij} e^Y_{ij} = \delta^{XY}$, and $e^X_{ii} = 0$. The index $X$ runs for the two dimensions of the transverse-traceless subspace of symmetric tensors. To avoid getting lost with the notation, we maintain the same symbols for the components in each basis. If $V_i^T$ is a transverse vector field, then $V_i^T = V_x^T e^x_i$, where $V_x^T$ are unconstrained quantities. Similarly, if $T^T_{ij}$ is a transverse-traceless tensor, $T^T_{ij} = T^T_X e^X_{ij}$. The introduction of these components yields a Jacobian in the measure of the path integral that depends only on the elements of the basis. Therefore, we discard this Jacobian. In Fourier space, the quadratic combinations of fields arising in the canonical Lagrangian can be written in terms of these components; for example $V_i^T V_i^T = V_x^T V_x^T$ and $T^T_{ij} T^T_{ij} = T^T_X T^T_X$. Since $V_x^T, T^T_X$ are unconstrained, the inverse can be computed directly in each basis.

We compute the inverse of the matrix of coefficients, keeping arbitrary values for the dimensionless constants coming from the conditions (\ref{gauechi}) -- (\ref{gaugechiiT}). Once the inverse has been obtained, we send $\epsilon \rightarrow 0$. Additionally, we find a set of values for the constants $\sigma_1,\sigma_2,\sigma_3$ that greatly simplifies the resulting propagators; specially for getting Klein-Gordon propagators in many places. These values are
\begin{equation}
 \sigma_1 = 0 \,, \quad \sigma_2 =1 \,, \quad \sigma_3 = \frac{1}{2} \,.
 \label{conditionssigma}
\end{equation}
With this choice, the complete form of the additional conditions we impose is 
\begin{equation}
	\begin{split}
		&
		h^L = - h^T \,,
		\\&
		\dot{n} = K^L + K^T  \,,
		\\&
		\dot{n}_i^T = \frac{1}{2} \Delta h_i^L  \,.
	\end{split}
\end{equation}

To present the resulting propagators, with $k^2 \equiv - \eta_{\mu\nu} k^\mu k^\nu =  \omega^2 - \vec{k}^2$, we define
\begin{equation}
\begin{split}
	&
	\Pi_0 = \frac{i}{k^2} \,,
    \\	&
	\Pi_1 = \frac{i}{k^2 - \mu } \,, 
	\\	&
	\Pi_{2} = \frac{i}{ k^2 + 4 \alpha^{2} \gamma\mu } \,.
\end{split}
\end{equation}
These factors have the form of Klein-Gordon propagators. They are valid in the frame defined by the ADM parametrization. Their poles give the quantum modes in the propagators, as we shall see. In order to have real solutions for the on-shell frequency $\omega$, we require $\mu \geq 0$. We impose $\kappa^{-2} \geq 0$ and $\alpha > 0$, such that we have a positive $\kappa^{-2}$. The factor $\Pi_2$ requires $\upsilon_3 < 0$ for the same reason.

We find that the quantum fields for which the poles can be identified in the cleanest way are the fields $h^T, K_x^L , h_X^T$. Here we focus on the propagators of these three fields, and in the appendix we present the full list of propagators of the BFV quantization. The diagonal propagators of the $h^T, K_x^L , h_X^T$ fields take the form 
\begin{equation}
\begin{split}
 &
 \langle h^{T}h^{T}\rangle =
 16i\gamma\upsilon_2 \Pi_1 \Pi_{2} \,,
 \\&
 - \vec{k}^{2} \langle K_x^L K_y^L \rangle = 
 \alpha^{-1} \Pi_{1} \delta_{xy} \,,
 \\& 
 \langle h_X^{T} h_Y^{T}\rangle 
 =  2 i \alpha^{-1} \Pi_{0} \Pi_{1} \delta_{XY} \,. 
\end{split}
\label{propagatorsfactorized}
\end{equation}

The $\kappa^{-2}=0$ limit (when general-relativity terms are decoupled) provokes a discontinuity in the way the individual modes are separated. This is related to the massless limit for all modes. Therefore, to further characterize the physics of the quantum modes, we must separate the totally massless case from the other case.

\paragraph{Case $\kappa^{-2} > 0$ : massless and massive poles.}
In this case, we can separate the propagators into partial fractions, each one with an order-one pole. We obtain
\begin{eqnarray}
    &&
	\langle h^{T}h^{T}\rangle=
	\frac{4}{3 \kappa^{-2}} \left( \Pi_2 - \Pi_1  \right)\,,
	\label{propagatorhscalar}
	\\&&
	- \vec{k}^{2} \langle K_x^{L} K_y^{L} \rangle
	= \alpha^{-1} \Pi_{1} \delta_{xy} \,,
	\\ &&
	\langle h_X^{T} h_Y^{T} \rangle 
	=  \kappa^{-2} \left( \Pi_0 - \Pi_1 \right) \delta_{XY} \,.
	\label{propagatorhtensor}
\end{eqnarray}
We have then two kinds of transverse-traceless tensorial poles: one massless ($\Pi_0$) and the other one massive ($\Pi_1$), the last one with squared mass equal to $\mu$. The first one represents the usual quantum modes of general relativity. The negative sign in front of $\Pi_1$ in (\ref{propagatorhtensor}) is associated with states of negative norm. Hence, this massive tensor is part of the inconsistent modes arising in the spectrum of this theory\footnote{We have fixed the signs of $\kappa^{-2}$ and $\alpha$. Other choices can translate the negative-norm states to other poles.}. We have also two kinds of scalars, both massive ($\Pi_1$ and $\Pi_2$), with squared masses $\mu$ and $4\alpha^2|\gamma|\mu$. Again, the negative sign in front of $\Pi_1$ in (\ref{propagatorhscalar}) indicates that the corresponding mode is physically an inconsistent mode. The vectorial sector has a unique massive mode ($\Pi_1$), with squared mass $\mu$, and with the right sign for the norm of the state. In total, these are the eight independent modes that this theory propagates.

\paragraph{Case $\kappa^{-2} = 0$ : totally massless poles.}
In the limit $\kappa^{-2} = 0$, the three factors become equal: $\Pi_0 = \Pi_1 = \Pi_2 = 1/k^2$, with a massless pole. From (\ref{propagatorsfactorized}), we see that in this limit the scalar sector $h^T$ acquires a massless pole of second order, the vectorial sector $K_x^L$ acquires a massless pole of order one, and the transverse-traceless tensorial sector $h_X^T$ acquires a massless pole of second order. Taking into account the double multiplicity of the poles of second order, these massless modes are the eight propagating modes of the theory in the $\kappa^{-2} = 0$ limit.

Another interesting sector is the set of BFV ghosts. By construction, these quantum fields are aimed to eliminate nonphysical degrees of freedom. We find that this sector is characterized by the massless $\Pi_0$ propagator. The representative of the scalar and vectorial sectors are, respectively,
\begin{eqnarray}
 &&
 \langle C\bar{C}\rangle = 
 - \Pi_0 \,,
 \label{propcc}
 \\&&
 	\langle C_x^{T}\bar{C}^{T}_y \rangle=
 - \Pi_0 \delta_{xy} \,, 
 \label{propcici}
\end{eqnarray}
whereas the propagator $\langle C^L \bar{C}^L \rangle$ is zero under the conditions we are using.

\section{Conclusions}
We have presented the BFV quantization of the quadratic gravity theory. It is based on the framework of the Hamiltonian formulation, which was developed by Buchbinder and Lyahovich \cite{Buchbinder:1987vp}. Instead of defining a quantum measure, the BFV quantization manages the first-class constraints in terms of auxiliary fields. This procedure has allowed us to obtain the full quantum canonical Lagrangian. Hence, the path integral we have obtained can be used directly for computations of diagrams, once it is expanded at the required order in perturbations. We have analyzed the gauge-fixing procedure from two perspectives: first, we have a mandatory condition necessary for the equivalence of the Hamiltonian formulation with the Lagrangian one. Second, we have studied the most general additional conditions based on very reasonable assumptions, as the features that they are local expressions, they do not introduce dimensional constants and they do not break the symmetry of spatial rotations. We have succeeded in making compatible both requirements, using the predefined form for fixing the additional conditions given in the BFV formalism. In this framework, there is a remaining BRST symmetry. We have computed the associated generator of this symmetry.

After introducing the additional conditions, we have computed the propagators of all quantum fields. Among them, there are propagators that lead to one-particle states with negative norm, in agreement with the results of Stelle in the Lagrangian approach \cite{Stelle:1976gc}. The masses of the poles depend on having active the general-relativity terms; that is, they are proportional to $\kappa^{-2}$. In this case, we have found five classes of modes: the massless transverse-traceless tensorial modes that one usually associates with the graviton of general relativity, a massive transverse-traceless tensorial mode, which is one of the negative-norm states, one massive transverse vector, and two massive scalars of different masses, one with positive norm and the other one with negative norm. It turns out that the negative-norm tensor and scalar modes have exactly the same mass as the vector mode, which has a positive norm. In the limit $\kappa^{-2} = 0$ all modes becomes massless.

The values of the masses we have found in this study equal the results of the spectrum found by Stelle in Ref.~\cite{Stelle:1976gc}, but distributed in a different way since we are using a different formalism for the quantum fields and a different (noncovariant) longitudinal-transverse decomposition. It would be interesting to delve deeper into this distribution of the problematic modes, since they arise in the tensor, vector, and scalar sectors, but with different signs of the propagators. For example, one could compute specific diagrams to study if there are some cancellations between the propagators of these modes.

An analysis that may continue this study is the problem of the integration on momenta, such that the (covariant) Lagrangian of the theory is obtained at the quantum level. This is very important for studying aspects such as the resulting quantum measure, a problem that was already pointed out by Buchbinder and Lyahovich \cite{Buchbinder:1987vp} using the Faddeev's scheme of quantization \cite{Faddeev:1969su}, and to check the validity of the standard Faddeev-Popov quantization that has been usually used in the quadratic gravity theory. This problem requires an appropriate use of the perturbative approach to implement the integration, since the explicit additional conditions are imposed in terms of the perturbative variables.

In implementation of the BFV quantization, we have used a specific additional condition, which is the vanishing of the trace of the perturbative spatial metric. As we have commented, this is mandatory to hold the equivalence  between the Hamiltonian and Lagrangian equations of motion. Several questions arise about this condition; specifically when one wonders about perturbations around different backgrounds, or at the nonperturbative level. The condition of the vanishing of the trace in perturbations around Minkowski spacetime was found computationally in Ref.~\cite{Bellorin:2025kir}. More analyses are required to elucidate whether this is a genuine feature of the complete, nonperturbative theory. Similarly, analysis of perturbations around other backgrounds may help us to understand this feature.

\appendix
\section{Propagators of the BFV quantization}
In this appendix, we present the full list of propagators for the quantum fields involved in the BFV quantization of quadratic gravity, under the conditions (\ref{conditionssigma}). We present only the nonzero diagonal propagators. 

The propagators of the scalar sector are characterized  by the $\Pi_1$ and $\Pi_2$ factors. The propagators are 
\begin{equation}
 \begin{array}{ll}
	\langle h^{T}h^{T}\rangle=
	16i\gamma\upsilon_2 \Pi_1 \Pi_{2}\,,
	&
	\langle \pi^{T}\pi^{T}\rangle =
	4i\gamma \mathcal{A}_1 \Pi_1 \Pi_2 \,,
	\\[1ex]
	\langle h^{L}h^{L}\rangle = 
	\langle h^{T}h^{T}\rangle \,,
	&
	\langle \pi^{L}\pi^{L}\rangle = 
	-i\gamma\kappa^{-4}\omega^{2}\mathcal{A}_{4}  \Pi_{0}^{2}\Pi_{1}\Pi_{2} \,,
	\\[1ex]
	\langle K^{T}K^{T}\rangle =
	4i\gamma\upsilon_2 \omega^2\Pi_1\Pi_{2}\,,
	&
	\langle p^{T}p^{T}\rangle =
	- 16i\gamma \mathcal{A}_{2} \Pi_{0}^{2}\Pi_{1}\Pi_{2} \,,
	\\[1ex]	
	\langle K^{L}K^{L}\rangle = 
	-i\gamma\omega^{2} \mathcal{A}_{5}\Pi_{0}^{2}\Pi_{1}\Pi_{2}
	\,,
	\quad
	&
    \langle p^{L}p^{L}\rangle = 
    16i\gamma v_{2}\kappa^{-4}\Pi_{1}\Pi_{2}
	\,, 
	\\[1ex]
	\langle nn\rangle = 
	- i\gamma \mathcal{A}_{3}\Pi_{0}^{2}\Pi_{1}\Pi_{2}
	\,,
	&
	-\vec{k}^2 \langle n^{L}n^{L}\rangle = 
	i\gamma {\displaystyle \frac{ \omega^{2} }{ \vec{k}^{2} } }   \mathcal{A}_{3} \Pi_{0}^{2}\Pi_{1}\Pi_{2} \,.
 \end{array}
\end{equation}
The propagators of the vectorial sector have in common the $\Pi_1$ factor:
\begin{equation} 
	\begin{array}{ll}
		- \vec{k}^{2} \langle h_x^{L} h_y^{L}\rangle
		= - 4 \alpha^{-1} \omega^2 \Pi_{0}^{2}\Pi_1 \delta_{xy} \,,
		\quad
		&
		- \vec{k}^{2} \langle \pi_x^{L} \pi_y^{L} \rangle
		= 4 \kappa^{-2}\mu \Pi_{1} \delta_{xy} \,,
		\\[1ex]
		- \vec{k}^{2} \langle K_x^{L} K_y^{L} \rangle
		= \alpha^{-1} \Pi_{1} \delta_{xy} \,,
		&
		- \vec{k}^{2} \langle p_x^{L} p_y^{L} \rangle
		= 4 \mathcal{A}_6 \Pi_{0}^{2}\Pi_{1} \delta_{xy} \,,
		\\[1ex] 
		\langle n_x^{T} n_y^{T} \rangle
		= \alpha^{-1} \vec{k}^{2} \Pi_{0}^{2}\Pi_{1} \delta_{xy} \,.
	\end{array}
\end{equation}
The propagators of transverse-traceless tensorial sector are characterized by the presence of the $\Pi_0$ and $\Pi_1$ factors. The propagators are
\begin{equation}
	\begin{array}{ll}
		\langle h_X^{T} h_Y^{T}\rangle 
		= 2i \alpha^{-1} \Pi_{0} \Pi_{1} \delta_{XY} \,,
		\quad
		&
		\langle \pi_X^{T} \pi_Y^{T}\rangle
		=  {\displaystyle \frac{i\kappa^{-2}}{2} } \mathcal{A}_7  \Pi_{0}\Pi_{1} \delta_{XY} \,, 
		\\[1ex]
		\langle K_X^{T} K_Y^{T} \rangle
		=  {\displaystyle \frac{i\omega^{2}}{2\alpha} } \Pi_{0}\Pi_{1} \delta_{XY} \,,
		&
		\langle p_X^{T} p_Y^{T}\rangle
		= -2i \kappa^{-2} \left( 3 k^{2} - 4\mu \right)\Pi_{0}\Pi_{1} \delta_{XY} \,.
	\end{array}
\end{equation}
For the ghost sector, the only nonzero propagators are those shown in (\ref{propcc}) -- (\ref{propcici}). In the previous expressions, we have introduced the combinations
\begin{eqnarray}
	\mathcal{A}_{1} &=& 
	- \omega^{2} \left( 4\alpha^{2} v_{1|4} \vec{k}^{4}
	+ 2 \kappa^{-2} \alpha v_{1|5} \vec{k}^{2}
	- \kappa^{-4}\upsilon_{3|8} \right)
	+ \alpha^{2} \left( \vec{k}^{2} + \mu \right)
	\left( 4\vec{k}^{2} - \mu \right)
	\left( 2 v_{1|3} \vec{k}^{2}
	- \kappa^{-2} \right) \,,
	\nonumber\\
	\mathcal{A}_{2} &=&
	2\alpha^{2} \upsilon_{1|3} \omega^{6} \left( 4\vec{k}^{2} - 5\mu \right) 
	- \omega^{4} \left( 4\alpha^{2} \upsilon_{5|16} \vec{k}^{4}
	- 2 \kappa^{-2} \alpha \upsilon_{9|29} \vec{k}^{2} 
	- \kappa^{-4} \upsilon_{11|32} \right) 
	\nonumber\\
	&&
	+2 \omega^{2} \vec{k}^{2} \left( 
	4\alpha^{2} \upsilon_{2|7} \vec{k}^{4}
	- \kappa^{-2} \alpha \upsilon_{3|13} \vec{k}^{2}
	- \kappa^{-4} \upsilon_{11|32}  \right) 
	\nonumber 
	\\&&
	- \vec{k}^{4} \left( 
	4\alpha^{2} \upsilon_{1|4} \vec{k}^{4}
	+ 2\kappa^{-2} \alpha \upsilon_{1|1} \vec{k}^{2}
	- \kappa^{-4} \upsilon_{11|32}  \right) \,,
	\nonumber\\
	\mathcal{A}_{3} &=&
	3\alpha \omega^{4}
	- \alpha \omega^{2} \left( 
	2\vec{k}^{2} + 3\mu \right) 
	+\vec{k}^{2} \left( \upsilon_{3|8} \vec{k}^{2} - \kappa^{-2} \right) \,,
	\nonumber\\
	\mathcal{A}_{4} &=&
	\upsilon_{11|32} \omega^{4}
	- \omega^{2} \left( 2\upsilon_{17|48} \vec{k}^{2}
	- 5 \kappa^{-2} \right)
	+ 9 \vec{k}^{2} \left(  
	\upsilon_{3|8} \vec{k}^{2} - \kappa^{-2} \right) \,,
	\nonumber\\
	\mathcal{A}_{5} &=&
	\upsilon_{3|8} \omega^{4}
	- \omega^{2} \left( 2 \upsilon_{5|16} \vec{k}^{2}
	- \kappa^{-2}  \right)
	+ \vec{k}^{2} \left( \upsilon_{11|32} \vec{k}^{2}
	- 5 \kappa^{-2} \right) \,,
	\nonumber\\
	\mathcal{A}_6 &=& 
	- 4\kappa^{-2}\mu \omega^{2}
	+ \kappa^{-2} 
	( 3\omega^{2} + \vec{k}^2 ) k^2 
	- \alpha  \vec{k}^{2} k^{4} \,,
	\nonumber \\
	\mathcal{A}_7 &=& 
	(3 \vec{k}^{2} + 4\mu) k^{2} + \mu \vec{k}^{2} \,,
	\nonumber \\
\end{eqnarray}
where
\begin{equation}
	\upsilon_{m|n} = m \alpha - n \beta 
\end{equation}
(such that $\upsilon_m = \upsilon_{1|m}$,  $\eta = \upsilon_{3|8}$ and $\tilde{\eta} = \upsilon_{5|16}$).



\end{document}